# Double glass transition in polyethylene naphthalate by MDSC, BDS, and TSDC

Juan Carlos Cañadas[1], José Antonio Diego[1], Sergio Diez-Berart[2], David Orencio López[2], Miguel Mudarra[1], Josep Salud[2], Jordi Sellarès*[1]

[1]DILAB, Departament de Física, ESEIAAT, Universitat Politècnica de Catalunya. Colom 11, 08222 Terrassa, Spain

[2]Grup de les Propietats Físiques del Materials (GRPFM), Departament de Física, ETSEIB, Universitat Politècnica de Catalunya. Diagonal, 647, 08028 Barcelona, Spain

**Corresponding Author**

Dr. Jordi Sellarès

Carrer de Colom 1, ESEIAAT, Edifici TR1, despatx 214

08222 Terrassa, Spain

Telf: +34 937398948

E-mail: jordi.sellares@upc.edu

**Abstract**

In this work, we present an experimental study of the primary and secondary relaxations of the semi-crystalline polymer polyethylene naphthalate (PEN), by Modulated Differential Scanning Calorimetry (MDSC), Thermally Stimulated Depolarization Currents (TSDC) and Broadband Dielectric Spectroscopy (BDS), and how they are affected by physical aging. Three dipolar relaxation modes can be observed: (from slowest to fastest) the primary α relaxation, which vitrifies at the glass transition temperature, $T_{g\alpha}$, and two secondary relaxations, named β* and β. MDSC results show how the secondary β* relaxation also vitrifies, giving rise to an additional glass transition at $T_{g\beta*}<T_{g\alpha}$. In fact, the α and β* relaxations can be considered as part of a very broad and distributed relaxation. Its main part is the primary α relaxation with a shoulder at the high-frequency region corresponding to a complex secondary β* relaxation. BDS results about the β* can be modeled by a main contribution ($\beta_3*$) and two additional ones ($\beta_1*$ and $\beta_2*$) with a weaker dielectric strength. TSDC results show that each single mode of the relaxation has its own glass transition temperature and they are compatible with the structure inferred by BDS. This scenario gives rise to an extended glass transition dually centered in the $T_{g\beta*}$~305 K and $T_{g\alpha}$~387 K temperatures.

## 1. Introduction

The glass transition to the non-equilibrium glassy state is attributed to the vitrification or "freezing" of the global molecular (or segmental, in polymers) reorientations when cooling the material down from an equilibrium disordered state [1-3]. When the vitrified material is heated up to the glass transition temperature ($T_g$), these frozen mechanisms get thermally activated and there is a jump in the heat capacity, $C_p$. The dielectric signature of the above-mentioned reorientations is the so-called α or primary relaxation mode. Actually, these movements are not rigorously frozen or stopped, but they are so slow that they last longer than the typical experimental time. As a criterion, when the characteristic time of this α relaxation mode is $\tau \sim 100$ s, it is said that the material reaches the glass transition temperature, as it is experimentally observed that such a temperature is approximately the same as the one where the jump in $C_p$ takes place.

One way to explain this behavior is via the free volume theory [4-6]. As the temperature goes down, the free volume ($V_f$) of the molecules (or the polymer chains) decreases, to the point that, at the glass transition temperature ($T_g$), the free volume is too low to permit molecular reorientations, and these are drastically slowed down. The material is in a non-equilibrium state and if it is cooled down further, the curve that relates free volume and temperature departs from the equilibrium line and becomes more horizontal, meaning that volume changes very slowly with temperature (Fig. 1). Once in the non-equilibrium glassy state, the material relaxes to the equilibrium state with time, usually when it is annealed at a constant temperature. This phenomenon is called physical aging [6-9]. During the aging process, the structural state of the material can be defined through the fictive temperature, $T_f$, which is the temperature at which the material would be at equilibrium, at the actual structural state [10,11]. This structural state can also be parameterized by the difference between the actual and the equilibrium values of volume or enthalpy [12-14]. A scheme of a typical aging experiment is presented in Fig. 1 [15].

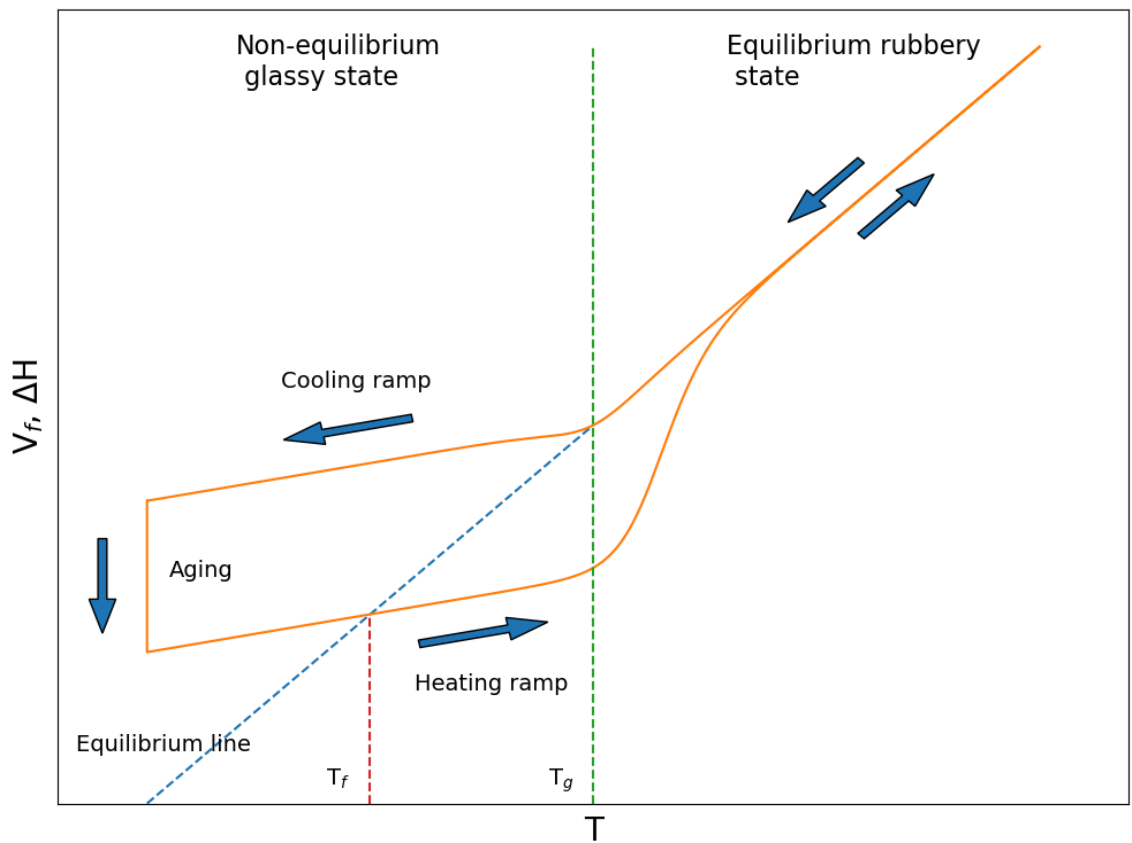

**Fig. 1.-** Scheme of enthalpy and free volume versus temperature in a physical aging

experiment for a glass forming material. If it is cooled fast enough $T_f$ is approximately $T_g$ during the cooling ramp for $T < T_g$. When the material is physically aged $T_f$ attains values lower than $T_g$. $T_f$ remains approximately constant during the heating ramp until $T$ approaches $T_g$, for fast enough heating ramps.

Together with the primary or α relaxation mode, glassy materials show additional relaxation modes at higher frequencies. The nature and behavior of these relaxations are subjects of intense research and debate, but they are usually assigned to: 1) splits from the primary α relaxation that are required to explain the viscosity of the materials when approaching $T_g$ (Johari-Goldstein relaxation [16-21]) and 2) local motions of small parts of the molecules [22,23]. All these relaxations remain active ($\tau$<100 s) below $T_g$ and it is not at all clear if they freeze at any temperature when the material is cooled further down. Even if they can reach values of $\tau$ larger than 100 s at some temperature, this fact does not seem to be accompanied by $C_p$ jumps at this same temperature. According to free volume models, if these relaxations are assigned to local or terminal movements of the whole molecules, such movements need a considerably lower volume than the one that is necessary for the α relaxation global or segmental reorientations, and as the free volume changes slightly after reaching the glass transition, it will never get low enough [6]. This would imply that secondary relaxations do not vitrify. As a consequence, the corresponding motions do not change while the material is under physical aging, which just affects the primary relaxation movements. In the case of a Johari-Goldstein secondary relaxation, it is considered a necessary precursor of the α relaxation and, therefore, it is intimately bound to the classical structural glass transition. Therefore, the existence of two glass transitions coming from the same disordered states of a glassy material is not usually considered.

Nevertheless, it has been shown for low-molar-mass non-symmetric odd liquid crystal (LC) dimers that two close glass transitions are present [24-26]. The disorder in LC phases is positional, as the molecules are orientationally ordered. Such materials are formed by two different elongated rigid cores, linked by a flexible alkyl chain with an odd number of methyl groups, and they present two relaxation modes in the disordered liquid crystalline phases when approaching the glass transition. It is shown that both modes vitrify at different temperatures: the slowest one, due to reorientations of the bulkier core, vitrifies at a temperature $T_{g1}$ and the faster one, due to the reorientations of the smaller core, vitrifies at $T_{g2}<T_{g1}$. Both kinds of reorientations have similar characteristic times, as both cores have comparable (yet different) volumes and masses and, therefore, their corresponding glass transition temperatures are just separated by a few degrees. The important fact is that both vitrify, but separately. Would it also be possible in other glassy materials? According to the free volume theory, it all depends on how much the volume necessary for the secondary relaxations can be lowered and if it can reach the critical value where the corresponding mechanisms get frozen. This possibility seems more feasible in systems where the difference in the characteristic times between the α relaxation and the secondary relaxation is lower. Besides, the α relaxation is usually much stronger than the secondary relaxations and, even if a secondary relaxation vitrifies, the corresponding jump in $C_p$ would be masked out by the jump related to the α relaxation. So, calorimetric data should be analyzed with extra care before discarding such a possibility.

We may cite the case of cyclooctanol, a glass-former Orientationally Disordered Crystal (ODIC) [27]. It shows two different disordered states, both of them vitrifying to their corresponding glassy states. The higher temperature phase (I) presents an α relaxation that vitrifies, and a secondary β relaxation, assigned to molecular conformational changes. The other phase (II) presents the same β relaxation, but not the α relaxation. There is evidence that this β relaxation, which is secondary in phase I, is the one that vitrifies in phase II.

Even in polymers, this possibility has been considered in studies about pharmaceutical drugs such as the polymer poly(vinylpyrrolidone) [28] even though this idea has been met with some skepticism [29].

Let us imagine a glassy material with two relaxation modes: the primary α one, related to global molecular or segmental reorientations, and a secondary β relaxation related also to global or segmental, yet different, and faster, reorientations. The glass transition would arise once the reorientations corresponding to the α relaxation get frozen. At temperatures below $T_g$, but not too far from it, reorientations related to the β relaxation would occur freely. When cooling the material down further, these β movements could also get frozen, as their available volume is not much different from the one related to the α reorientations.

The semi-crystalline polymer PEN presents three dipolar relaxation modes [30,31], the primary α relaxation, and two faster relaxations, named β* and β. Differential scanning calorimetry (DSC) and thermally stimulated depolarization currents (TSDC) determine that the α relaxation of PEN vitrifies at $T_g$=393 K [30,31]. Also, TSDC and broadband dielectric spectroscopy (BDS) [32], show that the β* mode appears at the low temperature (high frequency) end of the α mode and that it is very wide. The closeness of the α and β* modes leads us to think that PEN is an optimal candidate for analyzing the sub-$T_g$ behavior of a secondary relaxation of a polymer in order to see if it does vitrify. In particular, we will search in the calorimetric data to identify the signature of additional glass transitions. We also plan to determine the structure of the β* relaxation and elucidate its relationship with the α relaxation. Finally, we want to investigate how the different behavior of the modes that make up these relaxations intertwine to give rise to the extended glass transition that is observed calorimetrically.

This paper is organized as follows: in the next section, we will describe the materials, the experimental setup, and the experiments we have performed. In the following section, we will present our results and discuss their implications. We will end with some concluding remarks.

## 2. Methods

For MDSC and TSDC experiments we have employed PEN samples made from molten commercial pellets supplied by Goodfellow quenched into a sheet of 0.5 mm. BDS measurements require very smooth surfaces and extra thinness and for this reason, the samples have been obtained from commercial biaxially stretched sheets (Goodfellow) with a thickness of 75 μm heated above the fusion temperature and subsequently quenched to increase the amorphous fraction.

MDSC measurements were made by heating the sample at 2 K·min⁻¹ from 200 K up to 400 K ($>T_g$). Previously, the sample had been quenched from 400 K, to erase physical aging and start the experiment in an equilibrium structural state. Temperature amplitude and oscillation period were ±0.5 K and 60 s. A sample mass of 15 mg was selected.

BDS measurements were performed with an Alpha impedance analyzer from Novocontrol, for frequencies ranging from $10^{-3}$ Hz to $10^6$ Hz. The cell consists of two gold-plated brass electrodes (diameter 5 mm), making a plane capacitor about 75 μm thick. The sample is held in a Novotherm forced air oven, and the temperature is controlled via a System Quatro from Novocontrol. Dielectric measurements were performed on cooling with stabilization at different temperature steps and temperature control of 20 mK.

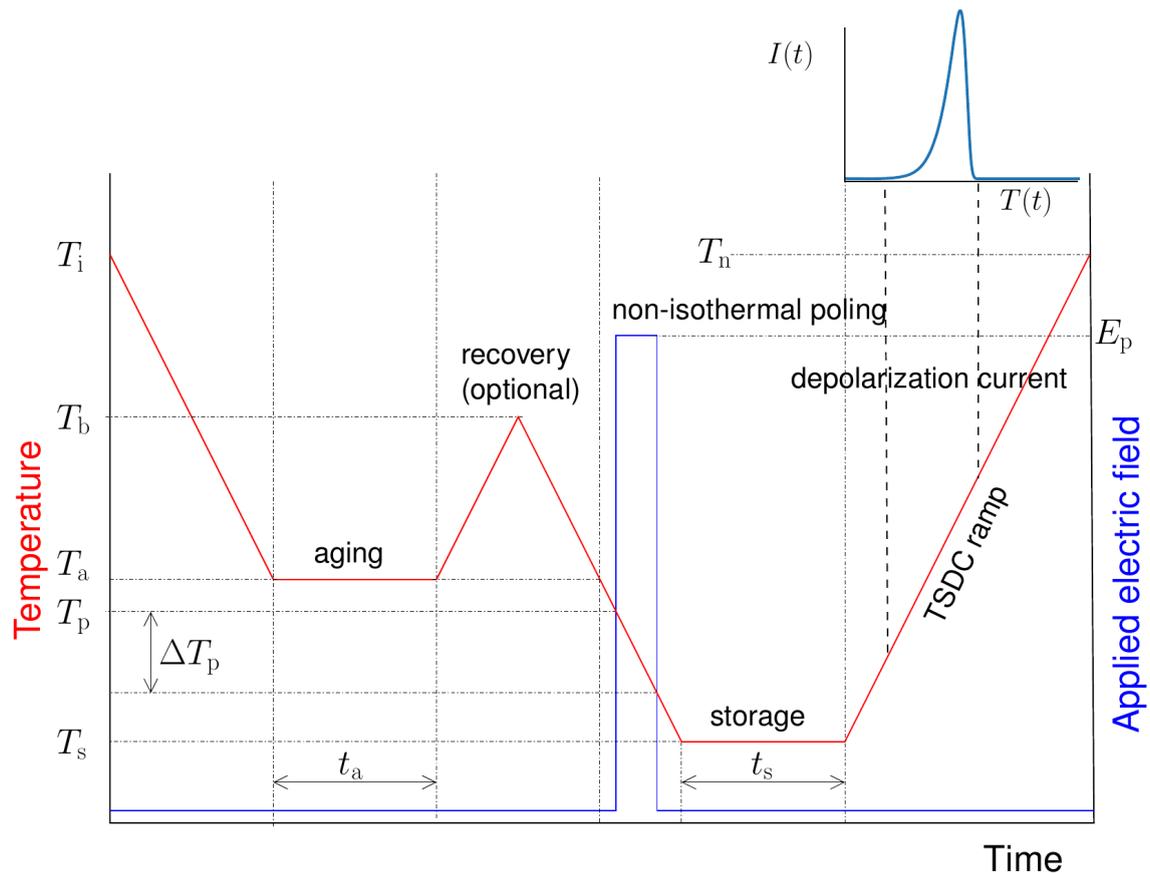

**Fig. 2.-** Scheme of thermal and electrical history in TSDC experiments. The recovery stage is optional. If this stage is not present the cooling ramp just goes from $T_a$ to $T_s$. See explanation in the text.

TSDC measurements were performed with an in-house assembled experimental setup that is described elsewhere [33]. The sample should begin in a well-defined structural state. This is accomplished by setting an initial temperature ($T_i$) above $T_g$. All the cooling and heating ramps of the experiment have a rate of 2 K/min. If the experiment involves aging, the sample is cooled down to an aging temperature ($T_a$) where it stays for an aging time ($t_a$). After aging, there can be an optional structural equilibrium recovery stage or the sample can be cooled down directly to the storage temperature ($T_s$). In either case, it is polarized non-isothermally during this cooling stage by an applied external field, starting at the poling temperature ($T_p$) and for a certain temperature window width ($\Delta T_p$). It is kept for a storage time ($t_s$) at a storage temperature ($T_s$) and, finally, it is heated up to a final temperature ($T_n$) while the thermally stimulated depolarization current is recorded. The procedure is schematized in Fig. 2.

Due to the low polarizability of PEN and, consequently, the low value of its depolarization current, the TSDC data presented in the next section have been passed either through a Butterworth low-pass filter or a running average filter in order to eliminate electrical noise. Special care has been taken to check that this does not alter the value of the currents in the region of interest.

## 3. Results and discussion

**Modulated Differential Scanning Calorimetry**

Fig. 3 shows the heat flow and reversing (static) $C_p$ of PEN obtained by MDSC from 220 K to 400 K. The heat capacity is determined as the heat flow divided by the heating rate and is formed by two contributions: reversing $C_p$, temperature dependent and consisting of the stored energy of active translational, rotational, and vibrational molecular modes; and non-reversing $C_p$, which is the kinetic response from any physical or chemical transformation [34]. In the case of glassy materials, as long as they experience a physical aging process below $T_g$, their enthalpy decreases and, when heated up to the glass transition temperature, there is an enthalpy recovery that is registered in the kinetic component of the heat capacity as an overshoot (Fig. 3). The primary glass transition is determined by the jump in $C_p$ that appears at $T_g$=387 K. An enthalpy overshoot is also recognizable in the heat flow. This is due to the non-isothermal structural recovery while the material is below the glass transition temperature.

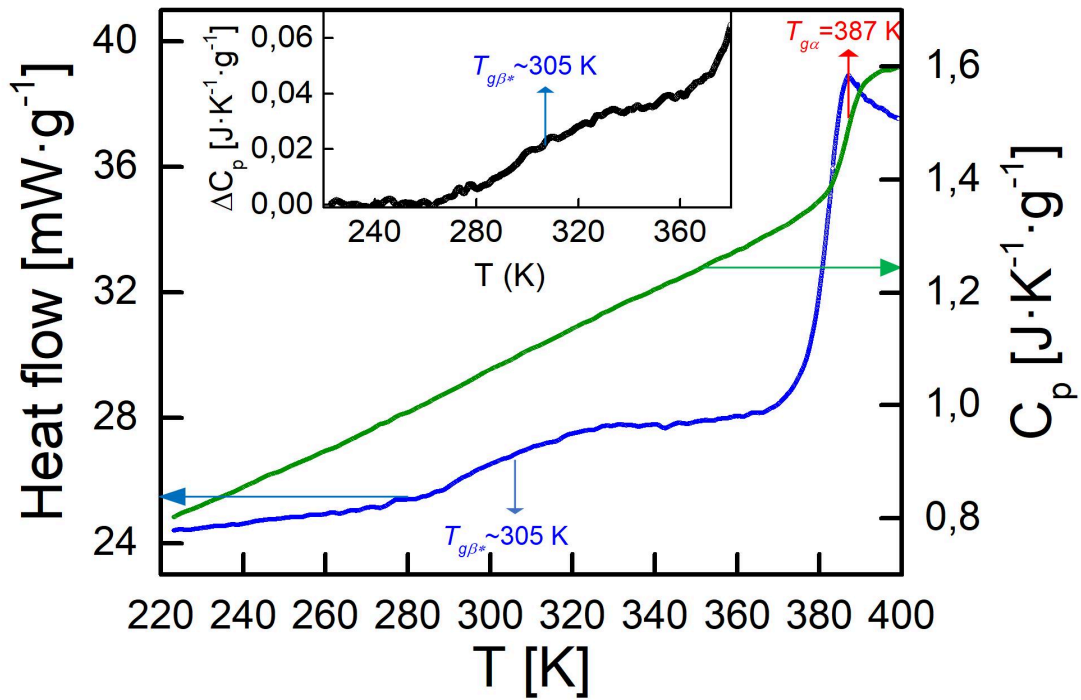

**Fig. 3.-** Heat flow (green line) and specific heat capacity (blue line) vs. temperature for PEN. The inset shows the difference between the specific heat capacity and the baseline of its values from 220 to 270 K, just before the extended glass transition related to the secondary β* relaxation mode takes place.

The heat flow shows a jump centered around 305 K. This jump is difficult to notice in the $C_p$, but if we draw a baseline parallel to the values of $C_p$ from 220 K to 270 K and subtract this line from the $C_p$ curve (inset of Fig. 3), a broad jump in $C_p$ is observed from ~270 K to ~370 K, just before the jump related to the primary glass transition. Such a jump can be divided into several steps, the largest of which is centered around 305 K. These results are evidence of the vitrification of a secondary relaxation, as the jump in $C_p$ indicates the activation of some previously frozen molecular modes. Unlike the primary relaxation, the secondary one is widely distributed, ranging from 270 to 370 K, and with a more pronounced contribution centered at $T_{g\beta*}$~305 K.

**Dielectric Spectroscopy**

We now present the dielectric behavior of the primary and secondary relaxation modes by means of BDS experiments. Fig. 4 shows the real and imaginary parts of the complex dielectric permittivity at several temperatures, where solid lines correspond to the fittings of experimental data. Each mode has been fitted according to the Havriliak-Negami function through the empirical relationship:

$$\varepsilon(\omega) = \sum_k \frac{\Delta\varepsilon_k}{\left[1+(i\omega\tau_k)^{\alpha_{HN,k}}\right]^{\beta_{HN,k}}} + \varepsilon_\infty - i\frac{\sigma_{DC}}{\omega\varepsilon_0} \qquad (1)$$

where k accounts for all the dielectric relaxation modes; $\Delta\varepsilon_k$ and $\tau_k$ are the dielectric strength and the characteristic time, respectively, of the k relaxation mode; $\alpha_{HN,k}$ and $\beta_{HN,k}$ are the shape parameters of the relaxation spectra; $\varepsilon_\infty$ is the dielectric permittivity at high frequencies and $\sigma_{DC}$ is the electric DC conductivity.

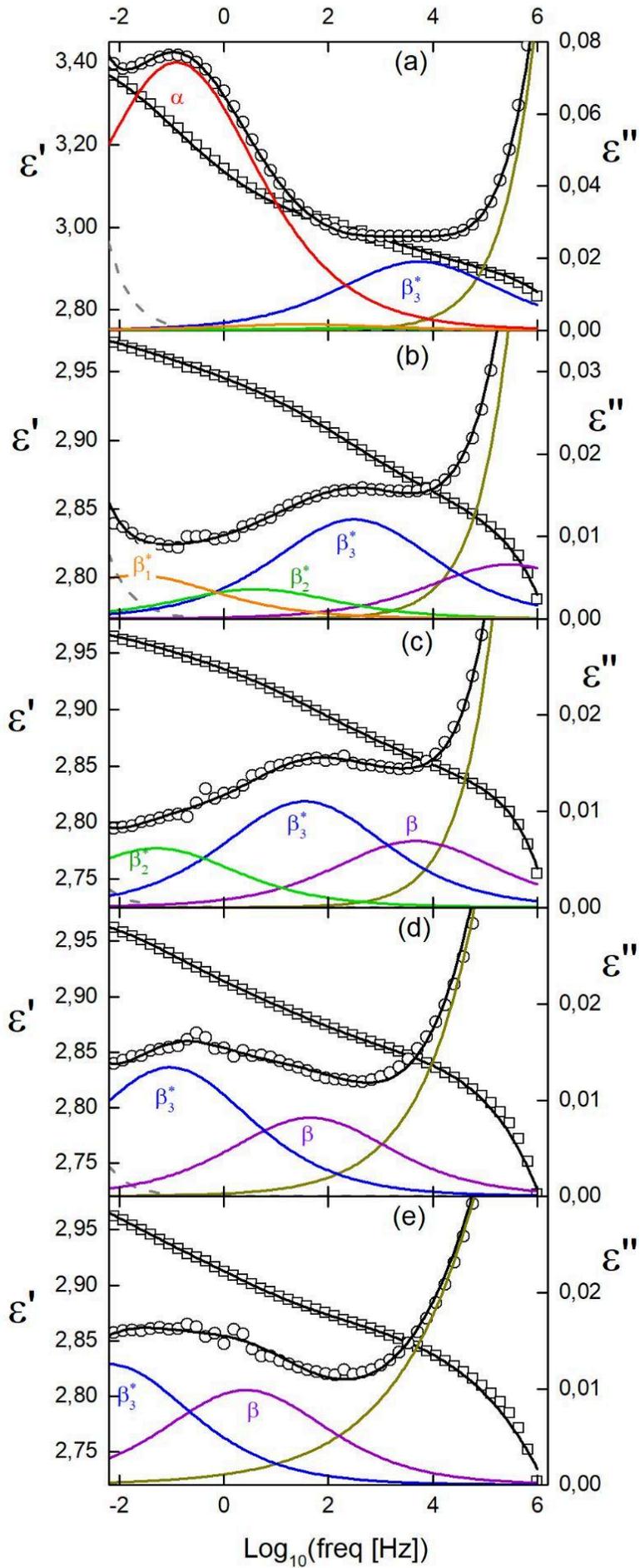

**Fig. 4**.- Frequency dependence of the complex dielectric permittivity for PEN at 400 K (a), 370 K (b), 350 K (c), 315 K(d), and 310 K (e). Squares account for the experimental real part, circles for the experimental imaginary part; fittings to Eq. 1 are shown by the solid lines: black solid lines show the whole dielectric response and colored solid lines represent the deconvoluted relaxation modes. Grey dashed lines account for DC conductivity.

The proper fitting of dielectric results to Eq. 1, requires the presence of at least six dielectric relaxation modes, which can be seen in Fig. 4. At 400 K (Fig. 4a), the primary α relaxation can be observed at about 0.1 Hz, together with a weaker secondary relaxation at ~10 kHz. We will call this mode $β_3^*$, a name that will be justified later. At frequencies above 1 MHz, an additional mode is present. Such a mode seems to be the superposed contribution of other secondary, tertiary... modes and of possible contact resistance. At 370 K (Fig. 4b), the α relaxation has left the frequency window as it is below 1 mHz. The $β_3^*$ mode is between $10^2$-$10^3$ Hz. Three more modes can be identified: One at $10^{-2}$-$10^{-1}$ Hz ($β_1^*$), another one, that is quite masked by $β_3^*$ (but which is clear at lower temperatures) at 1-10 Hz ($β_2^*$), and a third one at ~1 MHz (β), which is masked by the high frequency (>1 MHz) sum of contributions, but is also clearly separated at lower temperatures. At 350 K (Fig. 4c), the $β_1^*$ mode is below 1 mHz. $β_2^*$ (~0.1 Hz), $β^*_3$ (10-100 Hz), and β (~10 kHz), have moved to lower frequencies. Finally, at 315 K (Fig. 4d) and 310 K (Fig. 4e), we just can observe the $β_3^*$ (~0.1 Hz at 315 K and ~0.01 Hz at 310 K) and β (~100 Hz at 315 K and 1-10 Hz at 310 K) modes, together with the one at high frequencies.

We have identified and named the relaxation modes in accordance with previous results [30-32]. The primary α relaxation is quite clear. The very wide secondary β* relaxation is distributed in three more elementary modes: $β_1^*$, $β_2^*$, and $β_3^*$, this last one being the dominant in strength. The next mode in increasing frequency is the so-called β mode. All these five modes have been fitted with the parameters $α_{HN}$=0.4 and $β_{HN}$=1, meaning that they are Cole-Cole relaxations. Our frequency window just shows part of the additional mode at high frequencies (>1 MHz). We have performed the fitting of such a mode as a Cole-Cole one ($α_{HN}$=0.7 and $β_{HN}$=1).

Fig. 5 shows the behavior of the characteristic times of the relaxation modes with temperature in an Arrhenius plot. As we have shown when discussing Fig. 4, the $β_1^*$, $β_2^*$, and β modes are masked out by the stronger α, $β_3^*$, and high frequency modes, at some temperatures. Therefore, the values of the characteristic times for these three weaker modes have been obtained by keeping the fitting parameters controlled over the whole temperature range in order to get the most realistic results. Nevertheless, the values of these modes far from their corresponding glass transition temperatures are just tentative. It can be seen that the α relaxation mode reaches the glass transition ($τ$~100 s) at ~390 K and the modes that are part of the secondary β* relaxation reach their corresponding glass transitions at ~365 K ($β_1^*$), ~340 K ($β_2^*$), and ~305 K ($β_3^*$). As the $β_3^*$ mode is the strongest and, therefore, the dominant one, of those accounting for the β* relaxation, these results agree well with those from MDSC, where the primary glass transition is at 387 K and the secondary glass transition is very extended, with a dominant contribution centered at ~305 K.

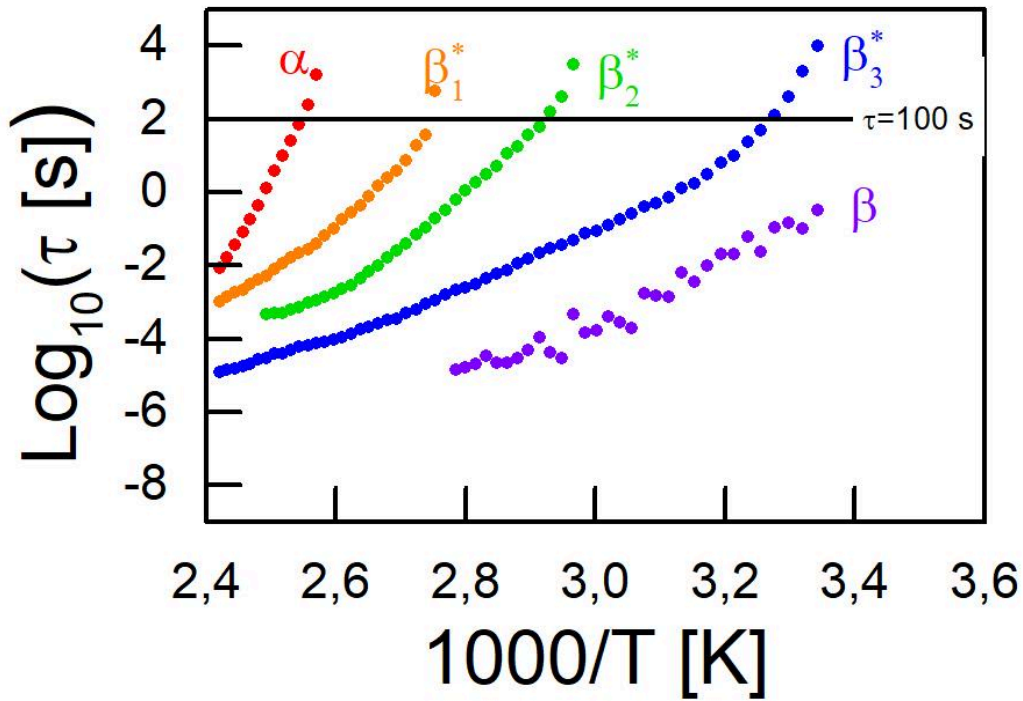

**Fig. 5.-** Arrhenius plot of the relaxation times of the different modes.

**Thermally Stimulated Depolarization Currents**

TSDC measurements allow for greater flexibility in a relaxational analysis since they are able to study specific parts of the relaxation by tailoring the poling process. Moreover, it is possible to study the material in different structural states if the sample is physically aged before the poling stage.

We will take advantage of this possibility to study how physical aging affects the relaxational spectrum. Whenever there is less free volume available segmental motions may become restricted, leading to reduced polarizability and dielectric strength.

In Fig. 6 we present the results obtained when the aging temperature ($T_a$) of 373 K is applied during different aging times ($t_a$) (indicated in the legend). Poling takes place after aging the sample, in the cooling ramp from aging temperature to storage temperature ($T_s$). The electric field is on through a wide range of temperatures ($T_p$ = 353 K and $\Delta T_p$ = 60 K) to obtain an image of the relaxational spectrum as complete as possible. Before the depolarization, the sample is kept for one hour ($t_s$) at 293 K ($T_s$) so that the TSDC current has a value close to 0 at the beginning of the TSDC ramp. This facilitates the interpretation of the relaxation, albeit at the expense of not capturing the modes that would relax near the beginning of the ramp.

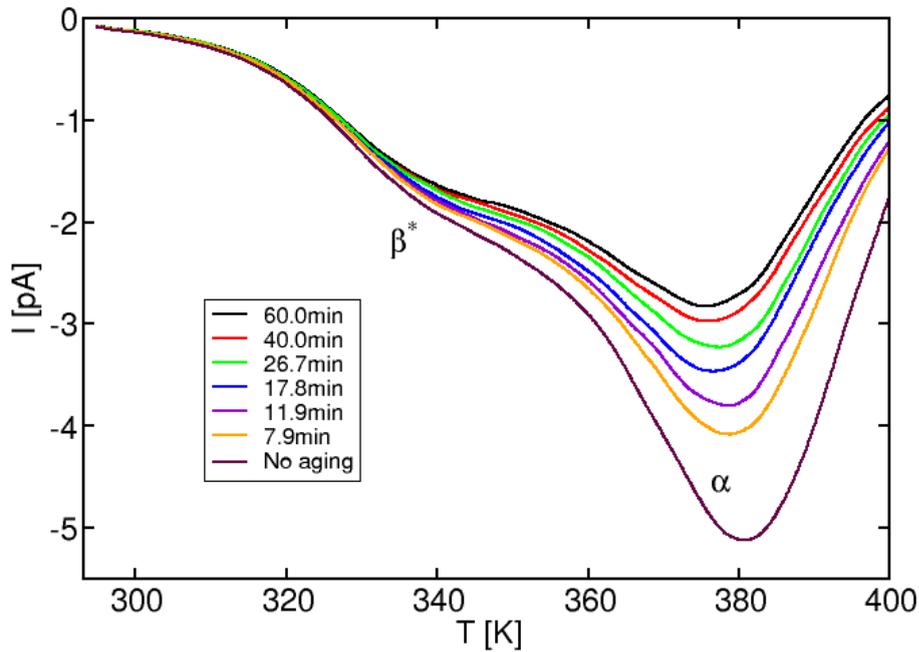

**Fig. 6.-** Depolarization currents for PEN when a wide non-isothermal poling window ($T_p = 353$ K and $\Delta T_p = 60$ K) is applied to the sample. Previous aging on the sample has been performed at $T_a = 373$ K for the time $t_s$ indicated in the legend.

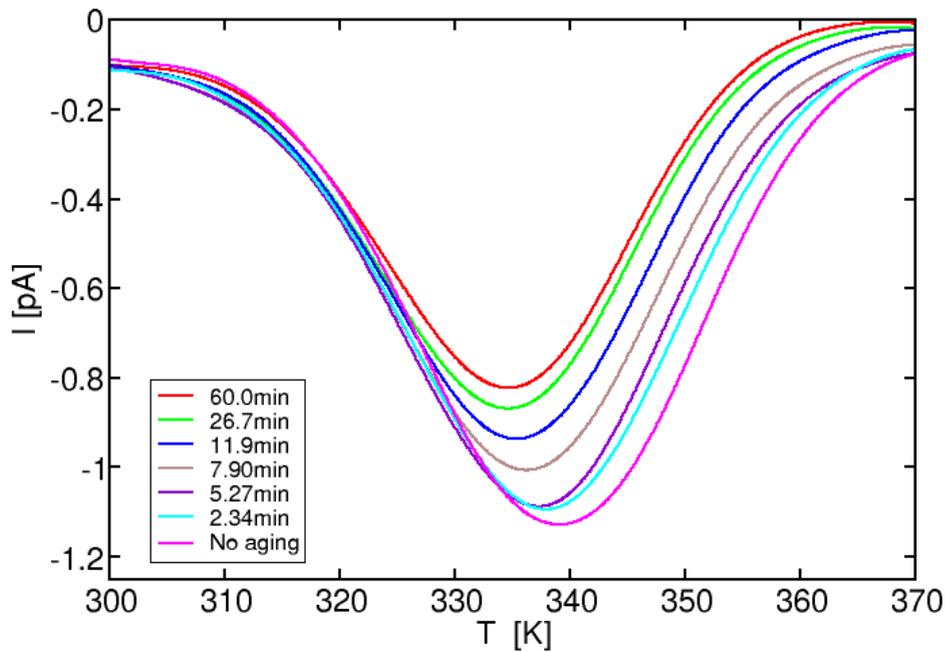

**Fig. 7.-** Depolarization currents for PEN when a narrow non-isothermal poling window ($T_p = 317$ K and $\Delta T_p = 22$ K) is applied to the sample. Previous aging on the sample has been performed at $T_a = 318$ K for the time $t_s$ indicated in the legend.

Results obtained confirm that the α peak is extended over a very broad range of temperatures, as can be seen in Fig. 6. We can identify a distinct shoulder attached to the α peak representing the β* relaxation [30-32], occurring at approximately 338 K. This observation suggests that the α relaxation exhibits a broad distribution of relaxation times (DRT) that encompasses the β* relaxation as a particular feature. This is evident as the shape of the depolarization current peak in a TSDC experiment closely depends on the DRT of the relaxation.

The temperature of the β* shoulder in Fig. 6 appears at the same temperature where we would expect to find the $β_2$* sub-relaxation identified by BDS. It is known that the temperature of the maximum of a depolarization current coincides with the glass transition temperature when the dielectric relaxation is related to a more general structural relaxation [35]. The β* shoulder in Fig. 6 is placed approximately at 340 K, which is the glass transition temperature of the $β_2$* sub-relaxation as obtained by BDS. On the contrary, the election of storage time and temperature is not suitable for studying the $β_3$* sub-relaxation because it should relax during the storage stage. The $β_1$* sub-relaxation is the weakest one and its peak should appear close to the main α peak. For these reasons, it can not be clearly appreciated in Fig. 6.

While the results presented in Fig. 6 clearly indicate that the α relaxation is affected by physical aging, there may be more uncertainty regarding the effect of aging on the β* relaxation. To elucidate if this relaxation also vitrifies, the results of another physical aging TSDC experiment are presented in Fig. 7. This experiment is similar to the previous one but this time the electric field is applied at $T_p$ = 317 K with $\Delta T_p$ = 22 K. Since this is a narrower range than before, only part of the relaxational spectrum is registered. In particular, the maxima of the current peaks of Fig. 7 are placed between 335 and 340 K, which is the range of the β* shoulder in Fig. 6 and is where the $β_2$* sub-relaxation studied by BDS should appear.

Also, in this experiment, the sample is aged at a lower temperature, $T_a$ = 318 K, before being poled. This temperature is about 70 K below $T_{gα}$ and therefore departure from structural equilibrium should be almost independent of the aging time because structural relaxation at this aging temperature should be extremely slow and physical aging should occur almost exclusively at the part of the cooling ramp from $T_i$ to $T_a$ just below $T_{gα}$. Instead, we find a strong dependence between the dielectric strength (peak area) of the relaxation and the aging time. This suggest that not only is the β* peak influenced by physical aging, but also that the glass transition temperature of the modes recorded in Fig. 7 must be significantly lower than $T_{gα}$ (but larger than $T_a$).

Another relevant observation in Fig. 7 is that the rate of decrease of the peak area is not uniform, despite the aging times being chosen in a logarithmic scale. This complex behavior has been previously noted by some authors [7] and may suggest that several sub-relaxations are involved in the aging process.

Recovery of structural equilibrium is as interesting as physical aging. As depicted in Fig. 1, free volume and enthalpy increase their values during a heating ramp, even in the glassy state, if T>$T_f$. This recovery is very low at the beginning but takes place at a much faster rate once T>$T_g$. If we are registering a relaxation by TSDC and, after aging the sample, we raise its temperature temporarily up to a recovery temperature ($T_b$) we can expect that the dielectric strength (area of the peak) begin once $T_b$ exceeds $T_g$.

To study the recovery of the structural equilibrium of the material we have devised the following experiment. During the cooling ramp, the sample is aged at $T_a$ = 318 K for $t_a$ = 60 min. Then it is heated up to a recovery temperature ($T_b$) for 30 seconds to attain some structural equilibrium recovery. Next, we activate the β* relaxation using a non-isothermal poling window with $T_p$ = 317 K and $\Delta T_p$ = 22 K. Finally, we obtain the depolarization current. Fig. 8 presents curves obtained using different recovery temperatures ($T_b$) as well as a curve without recovery stage and another curve with no aging at all, which should be considered as limit cases.

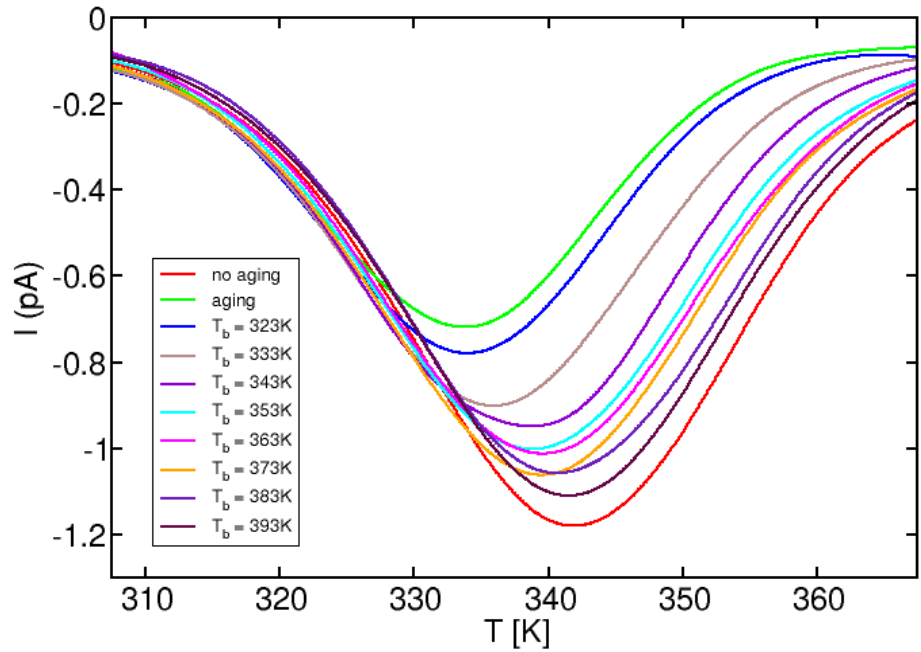

**Fig. 8.-** TSDC intensity corresponding to the β* relaxation with no aging, 60 min aging at $T_a$ = 318K with no recovery stage, and with recovery up to temperature $T_b$ as indicated in the legend.

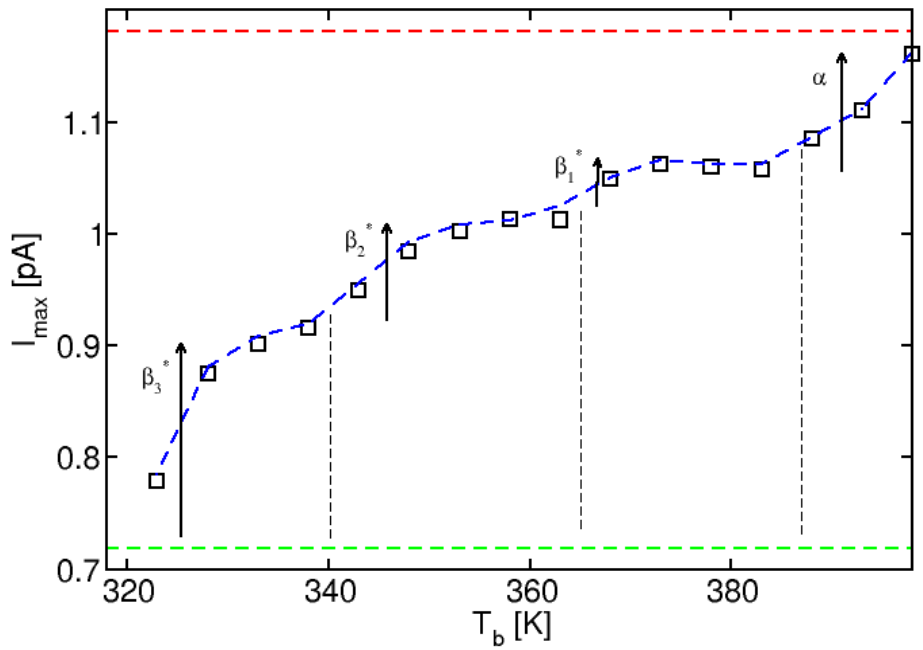

**Fig. 9.-** Maximum intensity ($I_{max}$) *vs.* $T_b$. Dashed horizontal lines represent limit cases (aging without heating after, and no aging). Arrows indicate particularly significant step-like increases in $I_{max}$. Vertical dashed lines are placed at 340 K ($T_g$ of $\beta_2^*$ by BDS), 365 K ($T_g$ of $\beta_1^*$ by BDS), and 387 K ($T_g$ of $\alpha$ by MDSC). The dashed line that runs between the points is just a guide for the eye.

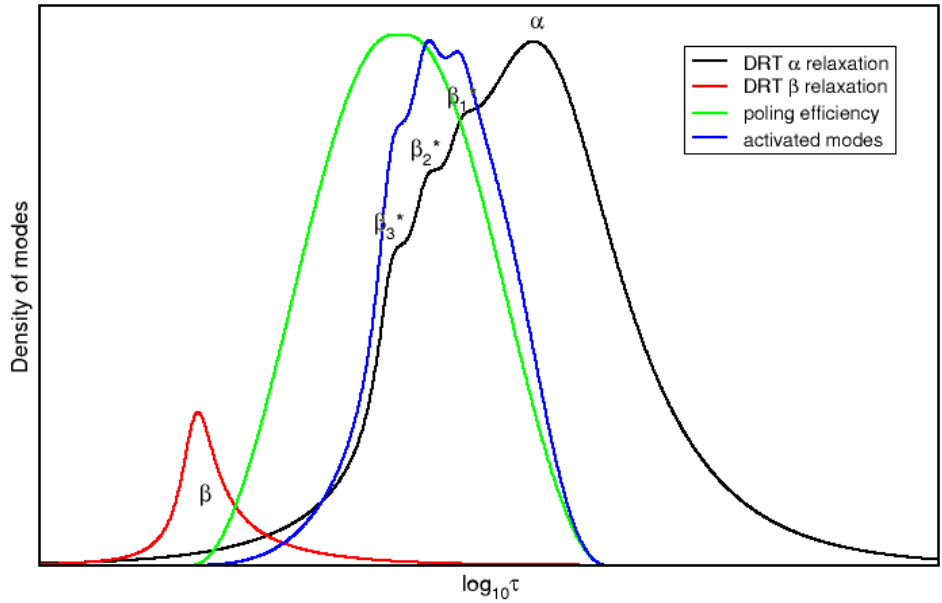

**Fig. 10.-** Qualitative interpretation scheme of the DRT of the dipolar relaxations, the poling efficiency, and the activated modes in narrow poling window TSDC experiments.

The first thing that should be noted is that the increase in peak area is gradual. This can only be interpreted as each of the many elementary modes composing the α and β* relaxations (described by the DRT) having its own $T_g$ and, therefore, they recover structural equilibrium at different temperatures. So, for a given $T_b$ only part of the modes (those with higher frequencies) recover from physical aging while the other ones remain frozen. It also implies that the poling parameters are not selective enough to activate an elementary relaxation and, therefore, there are several relaxations and sub-relaxations involved in the response of the sample.

The structural state can be tracked qualitatively by the area of the peak or the maximum intensity of the current. Both quantities should be roughly equivalent to the dielectric strength of the relaxation. We have plotted the maximum intensity of the current in front of $T_b$ in Fig. 9 to obtain a qualitative insight into the recovery of structural equilibrium. In this figure, it can be seen that this process is not uniform. We can observe that the increase in the recovery of the current intensity takes place in steps, marked with arrows in the figure. As mentioned earlier, a recovery temperature at which the dielectric strength begins to increase should match the $T_g$ of a sub-relaxation.

Effectively, we can observe in Fig. 9 that the steps at which the maximum intensity increases begin approximately at 340, 365, and 387 K which are the same $T_g$ temperatures estimated by BDS for the $β_2^*$ and the $β_1^*$ sub-relaxations and by MDSC for the α relaxation, respectively. On the other hand, the large increase between 320 and 335 K can be attributed to the low-frequency tail of $β_3^*$ (the part that can vitrify when the aging temperature is 318 K). Taken as a whole, results from the analysis of these data fit properly with our calorimetric and BDS results.

From this, it follows that the non-uniformity in the recovery of structural equilibrium is related to the structure of the DRT of the α and β* relaxations. Fig. 10 depicts our interpretation of the DRT of the dipolar relaxations. The DRT of α is not a clean bell-shaped function but rather has some shoulders on the high-frequency side (α+β*). We have identified three of these shoulders but an even more complex structure cannot be discarded. The main shoulder in the β* region of the DRT contributes to $C_p$ at 305 K whereas the maximum of the DRT (the α peak) contributes at 387 K, giving rise to the two different glass transitions detected by MDSC.

BDS is not able to discriminate between these sub-relaxations and all of them must be taken into account in the fitting process of the dielectric data. On its part, in TSDC we have more choice on the part of the relaxational spectra that we measure. Choosing appropriate parameters, a poling efficiency that activates mainly one zone of the DRT can be obtained, but even in this case, a mix of several relaxations and sub-relaxations is activated in our experiment because of their proximity. This situation is represented by the "poling efficiency" and "activated modes" curves in Fig. 10.

## 4. Conclusions

We have shown by means of MDSC that the glassy behavior of PEN is structured around an extended and continuous glass transition, with two main contributions at 387 K for the primary relaxation and 305 K for the secondary relaxation.

Moreover, using BDS and TSDC we have shown unequivocally that the dielectric β* relaxation vitrifies. The relaxational analysis of the dielectric data confirms that the MDSC secondary relaxation is the same as the dielectric β* relaxation.

Analysis by BDS reveals a complex structure for the β* relaxation. At least three β* sub-relaxations are needed to fit satisfactorily the experimental results: $β_1^*$, $β_2^*$, and $β_3^*$ (from lower to higher frequencies). The β* relaxation that has been studied by TSDC appears where the $β_2^*$ sub-relaxation observed by BDS should be expected, while the $β_3^*$ sub-relaxation is the most prominent in MDSC data.

The fact that both relaxations, α and β*, vitrify, suggests that they are related to segmental molecular motions. According to this image, the β* relaxation can be considered as a particular set of modes in a broader range that embraces the whole segmental motion space.

The study of the recovery of structural equilibrium by TSDC has proved to be a very insightful procedure. It shows good agreement with BDS results. On top of that, it shows that each one of the modes in which we can decompose the β* relaxation vitrifies at its own $T_g$. In essence, the emergence of more or fewer sub-relaxations, each with its own $T_g$, is a consequence of the distribution of relaxation times (DRT) within the α and β* relaxations, reflecting their particular structure.

All in all, the combined contribution of all the α and β* modes give rise to the observed extended double glass transition. This particular shape is due to the preeminence of the α relaxation and, to a lesser extent, the $β_3^*$ sub-relaxation.

Further research is needed to confirm the exact structure of the β* relaxation in PEN as well as to determine the segmental motions that cause it. For the moment, it seems clear that the α relaxation and related modes in PEN are more complex than in other and better studied polyesters such as PET.

CONFLICT OF INTEREST STATEMENT

The authors have no conflicts to disclose

DATA AVAILABILITY STATEMENT

The data that support the findings of this study are available from the corresponding author upon reasonable request.

ACKNOWLEDGEMENTS

The authors are grateful to R. Levit, A. Muñoz-Duque, and J. Sala-Vilaseca for technical support with the experiments.

This work has been funded entirely by our institution (UPC) through the allocation of ordinary research resources.

REFERENCES

[1] W. Kauzmann, Chem. Rev. **43**, 219 (1948).

[2] G. Adam and J. H. Gibbs, J. Chem. Phys. **43**, 139 (1965).

[3] J. Wong and C.A. Angell, *Glass: Structure by Spectroscopy*, Marcel Dekker, New York, 1976.

[4] F. Simon, Z. Anorg. Allgem. Chem. **203**, 219 (1931).

[5] A. J. Kovacs, J. Polym. Sci. **30**, 131 (1958)

[6] Struik, L. C. E. *Physical Aging in Amorphous Polymers and Other Materials*, Elsevier: Amsterdam, 1978.

[7] M. L. Cerrada and G. B. McKenna Macromol. **33**, 3065 (2000).

[8] K. Cheng and K. S. Schweizer, Phys. Rev. Lett. **98**, 167802 (2007).

[9] G. B. McKenna and S. L. Simon, Macromol. **50**, 6333 (2017).

[10] A. Q. Tool and C. G. Eichlin, J. Am. Ceram. Soc. **14**, 276 (1931).

[11] A. Q. Tool, J. Am. Ceram. Soc. **29**, 240 (1946).

[12] A. J. Kovacs, J. J. Aklonis, J. M. Hutchinson and A. R. Ramos, J. Polym. Sci., Part B: Polym. Phys. **17** 1097 (1979).

[13] D. Cangialosi, V. M. Boucher, A. Alegria and J. Colmenero, Phys. Rev. Lett. **111**, 095701 (2013).

[14] D. Cangialosi, *Vitrification and Physical Aging in Polymer Glasses by Broadband Dielectric Spectroscopy* in *Broadband Dielectric Spectroscopy: A Modern Analytical Technique*, American Chemical Society, Washington DC, 2021.

[15] J. Bayard, E. Dargent and J. Grenet, IEE Proc.-Sci. Meas. Technol. **145**, 53 (1998).

[16] M. Goldstein, J. Chem. Phys. **51**, 3728 (1969).

[17] G. P. Johari and M. Goldstein, J. Chem. Phys. **53**, 2372 (1970).

[18] K. L. Ngai, J. Non-Cryst. Sol. **275** 7 (2000).

[19] R. Brand, P. Lunkenheimer and A. Loidl, J. Chem. Phys. **116**, 10386 (2002).

[20] M. Paluch, C. M. Roland, S. Pawlus, J. Ziolo and K. L. Ngai, Phys. Rev. Lett. **91**, 115701 (2003).

[21] K. Geirhos, P. Lunkenheimer and A. Loidl, Phys. Rev. Lett. **120**, 085705 (2018).

[22] G. Williams and D. C. Watts, Trans. Faraday Soc. **67**, 1971 (1971).

[23] U. Schneider, R. Brand, P. Lunkenheimer and A. Loidl, Phys. Rev. Lett. **84**, 5560 (2000).

[24] S. Diez-Berart, D. O. López, J. Salud, J. A. Diego, J. Sellarès, B. Robles-Hernández, M. R. de la Fuente and M. B. Ros Materials **8**, 3334 (2015).

[25] J. A. Diego, J. Sellarès, S. Diez-Berart, J. Salud, J. C. Cañadas, M. Mudarra, D. O. López, M. R. de la Fuente and M. B. Ros, Liq. Cryst. **44**, 1007 (2017).

[26] D. O. López, J. Salud, M. R. de la Fuente, N. Sebastián and S. Diez-Berart, Phys. Rev. E **97**, 012704 (2018).

[27] R. Puertas, M. A. Rute, J. Salud, D. O. López, S. Diez, J. C. Van Miltenburg, L. C. Pardo, J. Ll. Tamarit, M. Barrio, M. A. Pérez-Jubindo and M. R. de la Fuente, Phys. Rev. B **69**, 224202 (2004).

[28] S. Vyazovkin and I. Dranca, Pharm. Res. **23** (2), 422 (2006).

[29] E. O. Kissi, H. Grohganz, K. Löbmann, M. T. Ruggiero, J. A. Zeitler and T. Rades, J. Phys. Chem. B **122**, 2803 (2018).

[30] T. A. Ezquerra, E.I. Balta-Calleja and H.G. Zachnann, Acta Polymer. **44**, 18 (1993).

[31] J. C. Cañadas, J. A. Diego, J. Sellarès, M. Mudarra, J. Belana, R. Díaz-Calleja and M. J. Sanchis, Polymer **41**, 2899 (2000).

[32] L. Hardy, A. Fritz, I. Stevenson, G. Boiteux, G. Seytre and A. Schönhals, J. Non-Cryst. Sol. **305**, 174 (2002).

[33] J. C. Cañadas, J. A. Diego, M. Mudarra, S. Parsa and J. Sellarès, J. Phys. D **52**, 155301 (2019).

[34] K. J. Jones, I. Kinshott, M. Reading, A. A. Lacey, C. Nikolopoulos and H. M. Pollock, Thermochimica Acta **304-305**, 187 (1997).

[35] J. Belana, P. Colomer, S. Montserrat and M. Pujal, J. Macromol. Sci. B **23**, 467 (1984).